\documentclass[prx,letterpaper,aps,floatfix,superscriptaddress,twocolumn]{revtex4-1}
\usepackage{graphicx}
\usepackage{amsmath}
\usepackage{amsfonts}
\usepackage{enumitem}
\usepackage{subfigure}
\usepackage{color}
\newcommand{\beq}{\begin{equation}}
\newcommand{\eeq}{\end{equation}}
\newcommand{\bk}{{{\bf{k}}}}

\newcommand{\bQ}{{{\bf{Q}}}}
\newcommand{\br}{{{\bf{r}}}}

\newcommand{\bG}{{{\bf{G}}}}

\newcommand{\bq}{{\bf{q}}}

\newcommand{\beqa}{\begin{eqnarray}}
\newcommand{\eeqa}{\end{eqnarray}}
\newcommand{\pdg}{{\vphantom \dag}}
\newcommand{\dg}{{\dag}}
\newcommand{\bnabla}{{\boldsymbol \nabla}} 

\newcommand{\btau}{{\boldsymbol \tau}}

\newcommand{\cI}{{\cal I}}

\newcommand{\cD}{{\cal D}}

\newcommand{\bS}{{{\bf{S}}}}
\begin{document}
\title{Current-Induced Spin Accumulation and Magnetoresistance in Chiral Semimetals}
\author{A.A. Burkov}
\affiliation{Department of Physics and Astronomy, University of Waterloo, Waterloo, Ontario 
N2L 3G1, Canada} 
\affiliation{Perimeter Institute for Theoretical Physics, Waterloo, Ontario N2L 2Y5, Canada}
\author{Michael Smith}
\affiliation{Materials Science Division, Argonne National Laboratory, Lemont, Illinois 60439, USA}
\author{Alexander Hickey}
\affiliation{Department of Physics and Astronomy, University of Waterloo, Waterloo, Ontario 
N2L 3G1, Canada} 
\author{Ivar Martin}
\affiliation{Materials Science Division, Argonne National Laboratory, Lemont, Illinois 60439, USA}
\date{\today}
\begin{abstract}
Weyl fermions possess the property of spin-momentum locking: the expectation value of the spin is parallel or antiparallel to the momentum at any given 
point in the Brillouin zone in the vicinity of a Weyl node. 
This is a direct consequence of the fact that Weyl nodes are monopoles of the Berry curvature, and in this sense an expression of the nontrivial Weyl electronic structure topology. Thanks to this property, an isolated Weyl node produces a large spin accumulation in response to a charge current, $\hbar/2$ per 
electron, similar to surface states of time-reversal invariant topological insulators. However, in bulk Weyl semimetals, the nodes must occur in pairs 
of opposite chirality and, when the nodes are at the same energy, the effect cancels out. 
Here we show that this cancellation is avoided in chiral semimetals, in which Weyl nodes of opposite chirality occur at different energies due to broken mirror symmetry. 
We find that the spin accumulation is maximized when the Fermi energy coincides with one of the nodes in a pair and reaches 
the same value as for an isolated node in this case. Moreover, we demonstrate the existence of a distinct magnetoresistance mechanism,
closely related to this current-induced spin accumulation. 
\end{abstract}
\maketitle
\section{Introduction}
\label{sec:1}
Topological metals is a rapidly growing field of research, due to both fundamental and applied interest~\cite{Volovik03,Volovik07,Weyl_RMP,Murakami07,Wan11,Burkov11-1}. 
The standard view is that most of their ``topological" properties may be related to the chiral anomaly~\cite{Weyl_encyclopedia}, which 
leads, in particular, to the observed negative longitudinal magnetoresistance~\cite{Spivak12,Burkov_lmr_prb,Ong_anomaly,Li_anomaly}
and the extra magnetic-field-dependent peak in the low-frequency optical conductivity~\cite{Burkov_Drude,Armitage21}. 

A closely related, but physically distinct property of topological metals, that has not received as much attention, is spin-momentum locking~\cite{Mertig18,Zhang19,Haney21}. 
This results from the fact that Weyl nodes [we will focus on Weyl point-node topological semimetals in this article, with pairs of nondegenerate bands 
touching at discrete points in the first Brillouin zone (BZ)] act as sources and sinks, or monopoles, of the Berry curvature. 
As a result, a two-dimensional (2D) Fermi surface sheet, enclosing a single Weyl node, is characterized by a nonzero Chern number. 
This is analogous to a similar property in three-dimensional (3D) time-reversal (TR) invariant topological insulators (TI)~\cite{Hasan10,Qi11}. 
While the existence of the 2D Dirac surface states in this case may be discussed in terms of the parity anomaly and the topological theta-term in the action for the electromagnetic field, a related, but distinct, property of these states is a $\pi$ Berry phase, accumulated on a closed-loop path in momentum space, enclosing the surface Dirac point. This leads to spin-momentum locking and current-induced spin accumulation in the 3D TI surface states~\cite{Burkov10,Ralph14,Ando14,YongChen15,Jonker16}.

An isolated Weyl node is characterized by a strong spin-momentum locking, similar to the 2D Dirac surface state of a TR-invariant TI. 
The analogy becomes particularly transparent if one views a single Weyl cone as a surface state of a 4D quantum Hall insulator. 
However, in 3D Weyl semimetals the band-touching nodes always come in pairs of opposite chirality. If the nodes in such pairs have to appear at the same 
energy due to crystalline symmetry (inversion or mirror), the spin accumulation in response to a charge current cancels out between the two nodes in any given pair. On the other hand, if the inversion and mirror symmetries are broken and the opposite chirality nodes appear at different energies~\cite{Zyuzin12-2}, spin accumulation becomes possible~\cite{Moore17,Orenstein20,Nomura21,Law21}. 

In this paper, we develop a theory of current-induced spin accumulation in chiral topological semimetals, in which nodes of opposite chirality appear at 
different energies due to broken inversion and mirror symmetries. 
Using a simple toy model of a pair of opposite-chirality Weyl nodes, separated in both energy and momentum, and taking impurity scattering into account, 
we derive a set of coupled spin-charge diffusion equations, which describe the intertwined transport of nonequilibrium spin and charge densities 
in a generic chiral semimetal. 
We demonstrate that the spin accumulation in response to a charge current in such semimetals has a nontrivial nonmonotonic dependence on the Fermi energy. 
In particular, the maximum spin accumulation, $\hbar/2$ per electron, is achieved when the Fermi energy coincides with the location of 
one of the nodes and decays away from this value. 
In addition, we demonstrate a closely related to spin accumulation magnetoresistance effect, when spin injection from a ferromagnetic contact leads to an extra contribution to the voltage drop, which depends on the orientation of the magnetization of the ferromagnetic contact relative to the current. 
\section{Theory of current-induced spin accumulation in chiral semimetals}
\label{sec:2}
\begin{figure}
\centering
\includegraphics[width=\linewidth]{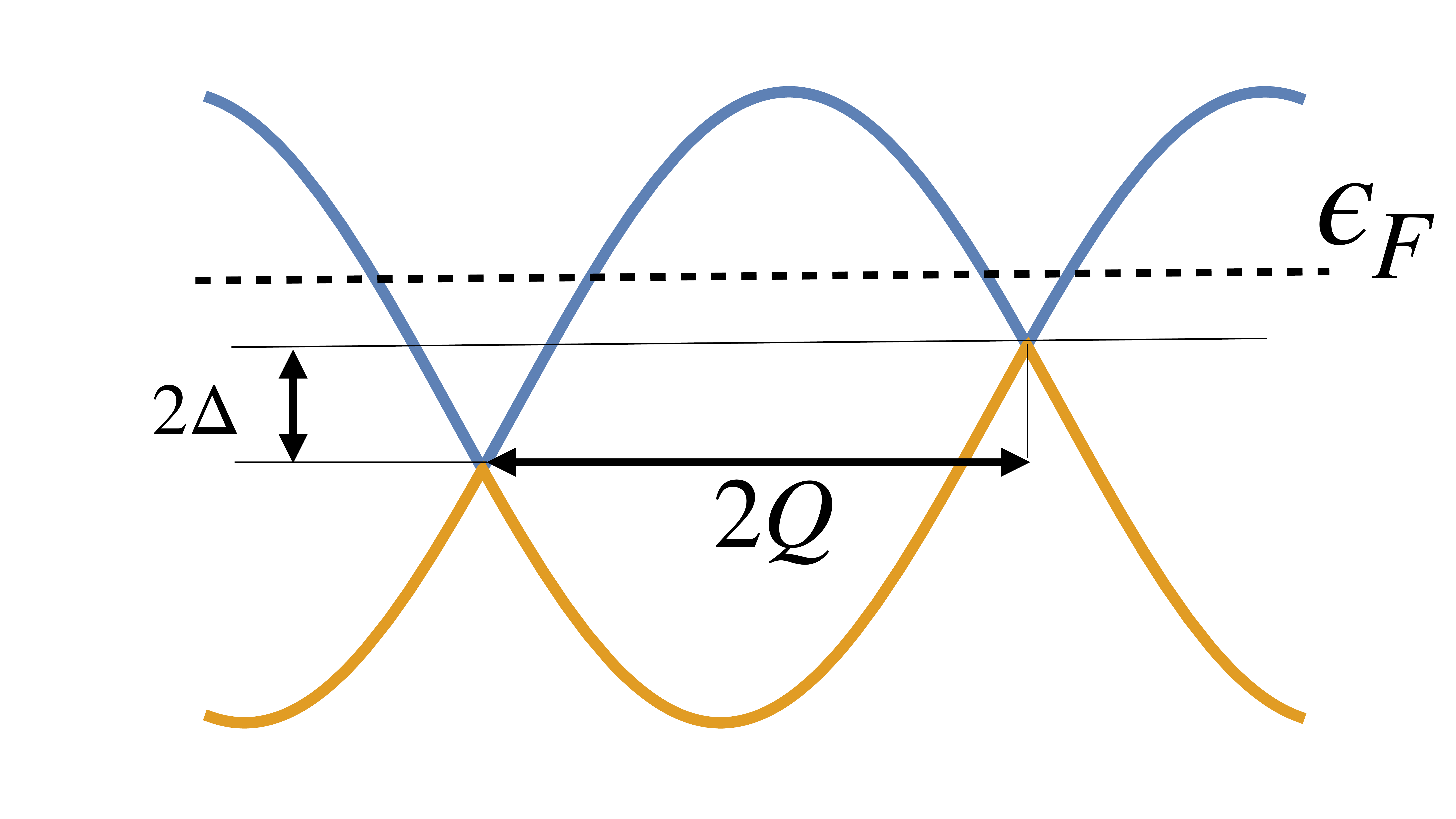}
\caption{(Color online) Schematic band structure of a chiral semimetal. A pair of nodes of opposite chirality are separated by an energy $2 \Delta$ and a wavevector $2 Q$.
Typical values of $\Delta$ in known chiral semimetals, such as CoSi, range from about ten to a few hundred meV.}
\label{fig:1}
\end{figure}

We start from the simplest toy model of a chiral topological semimetal, which consists of a pair of linearly-dispersing Weyl nodes, separated in momentum 
by $2 Q$ and in energy by $2 \Delta$ (see Fig.~\ref{fig:1})
\beq
\label{eq:1}
H = \sum_{\bk s}  c^\dg_{\bk s} \left[s v_F \btau \cdot (\bk - s \bQ) + s \Delta \right] c^\pdg_{\bk s}, 
\eeq
where $s = \pm$ labels the Weyl node chirality, $v_F$ is the Fermi velocity and $\hbar = 1$ units will be used throughout. 
$\btau$ in Eq.~\eqref{eq:1} describes the two bands, that touch at the Weyl nodes
and will always contain some admixture of orbital degrees of freedom in addition to pure spin~\cite{Burkov11-1}, but this is not important for our 
purposes, since it will only affect the results quantitively ($\btau$ has the same symmetry properties as the spin and is thus 
``proportional" to it, in the sense of the Wigner-Eckart theorem). 
Real currently known chiral semimetals, such as RhSi and CoSi, also have a significantly more complex electronic structure~\cite{Hasan17,Pshenay-Severin_2018,Orenstein20}, but the qualitative effects we describe here should not depend significantly on details. 

\begin{figure}
\centering
\includegraphics[width=\linewidth]{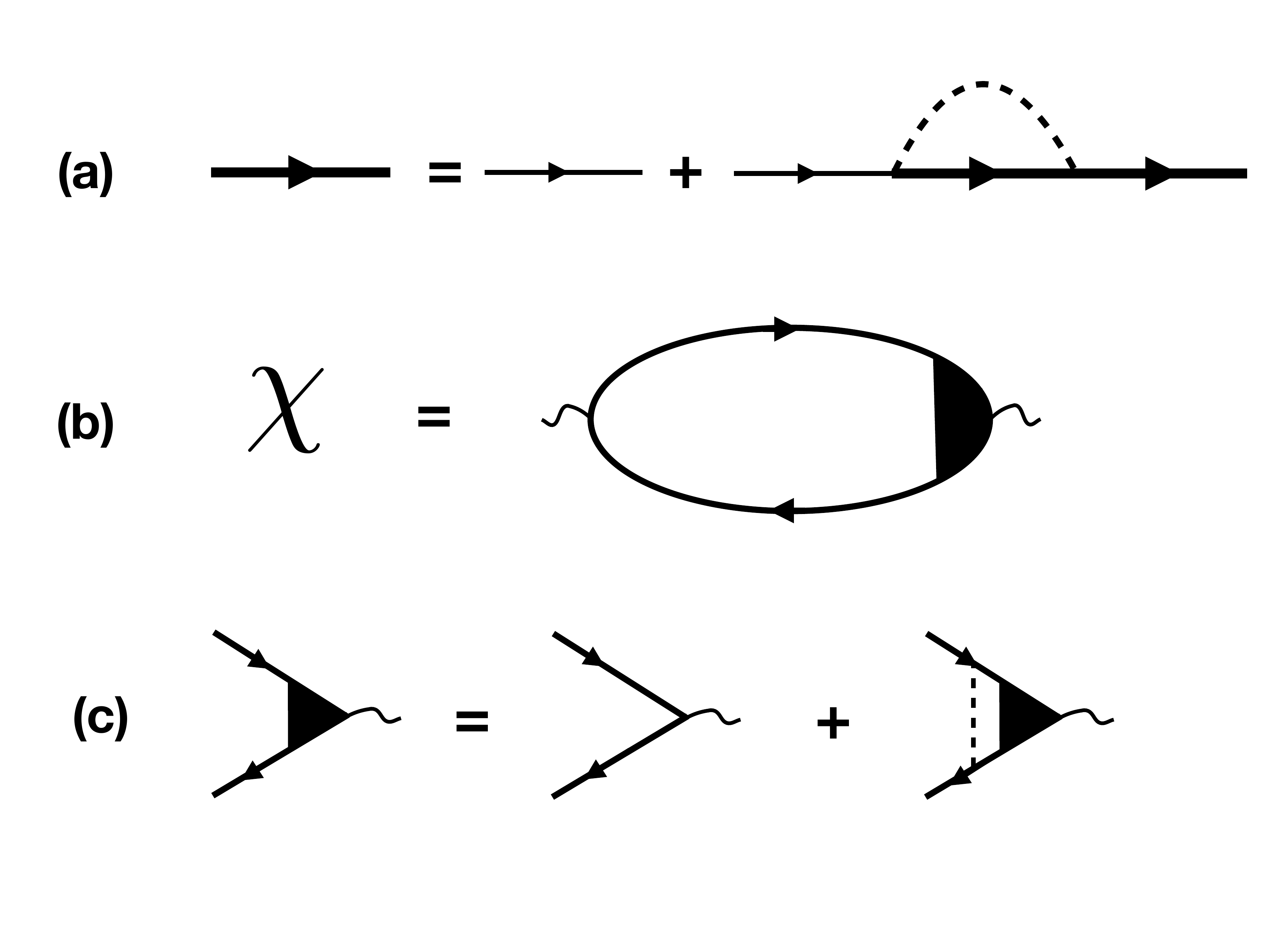}
\caption{Diagrammatic representation of (a) SCBA Green's function. Thin line represents the bare Green's function, thick line is the SCBA impurity-averaged Green's function and the dashed line represents the disorder potential correlator $\langle V(\br) V(\br') \rangle = \gamma^2 \delta(\br - \br')$. (b) Density response function $\chi$. (c) Equation for the diffusion vertex $\cD$.}
\label{fig:2}
\end{figure}

Since we will be interested in low-frequency transport phenomena on macroscopic scales, it is important to take account of the impurity scattering. 
We will do this using the standard diagrammatic perturbation theory for the charge and spin density response \cite{Altland10}. 
As is well-known, the simplest set of approximations, fully respecting conservation laws, involves self-consistent Born approximation (SCBA) for the 
impurity self-energy, or scattering rate, along with ladder vertex corrections, see Fig.~\ref{fig:2}. 
We take the impurity potential $V(\br)$ to have the white noise form for simplicity, such that $\langle V(\br) \rangle = 0$ and $\langle V(\br) V(\br') \rangle = \gamma^2 \delta(\br - \br')$, where the angular brackets denote averaging with respect to disorder realizations. 
The SCBA scattering rate in our case is independent of either momentum or band index and is given by $1/\tau = \pi \gamma^2 g$, where 
$g = (\epsilon_F^2 + \Delta^2)/ \pi^2 v_F^3$ is the density of states at the Fermi energy $\epsilon_F$. 

The impurity-averaged SCBA retarded Green's functions may be expanded in the basis of a unit matrix $\tau^0$ plus three Pauli matrices $\btau$ as
$G^R(\bk, s,  \omega) = G^R_{0}(\bk, s, \omega) \tau^0 + {\bf G}^R(\bk, s, \omega) \cdot \btau$, 
where 
\beqa
\label{eq:2}
G^R_{0}(\bk, s, \omega)&=&\frac{1}{2} \sum_{r} \frac{1}{\omega - \xi_{s r}(\bk) + \frac{i}{2 \tau}}, \nonumber \\
\bG^R(\bk, s, \omega)&=&\frac{s \bk}{2 k} \sum_{r} \frac{r}{\omega - \xi_{s r}(\bk) + \frac{i}{2 \tau}}.
\eeqa 
Here $\xi_{s r}(\bk) = r v_F |\bk - s \bQ| + s \Delta - \epsilon_F$ are the band eigenstate energies, counted from the Fermi energy, and $r = \pm$ labels 
the conduction and valence bands. 

To obtain the transport equations, we evaluate the retarded density matrix response function, defined as
\beqa
\label{eq:3}
&&\chi_{a b}(\br, t | \br', t') = - \frac{i}{2} \theta(t - t') \tau^a_{\sigma_2 \sigma_1} \tau^b_{\sigma_3 \sigma_4} \nonumber \\
&\times&\langle [ \varrho^\pdg_{\sigma_1 \sigma_2}(\br, t), \varrho^\dg_{\sigma_3 \sigma_4}(\br', t') ] \rangle, 
\eeqa
where $a, b = 0, x, y, z$, summation over spin indices is implicit and 
$\varrho_{\sigma_1 \sigma_2}(\br, t) = \sum_s \Psi^\dg_{\sigma_2}(\br, s, t) \Psi^\pdg_{\sigma_1}(\br, s, t)$ is the density matrix. 
This is evaluated using SCBA plus ladder vertex corrections, which is an accurate approximation as long as the quasiparticle energy in the vicinity of each 
Weyl node exceeds the scattering rate. Strictly speaking, this breaks down whenever the Fermi energy approaches either one of the nodes, 
i.e. $\epsilon_F \approx \pm \Delta$, since the scattering rate always stays finite due to internode scattering. 
However, since all of the results, as will be seen below, are smooth functions of $\epsilon_F$, we can be fairly confident this approximation is at 
least qualitatively correct for any value of the Fermi energy. Physically, it amounts to neglecting quantum interference processes, which should always 
be justified in 3D when disorder is not too strong and temperature is not too low. 

In this case, one obtains 
\beqa
\label{eq:4}
\chi_{a b}(\bq, \omega)&=&\int_{-\infty}^{\infty} \frac{d \epsilon}{2 \pi i} \left[\omega \frac{d n_F(\epsilon)}{d \epsilon} 
P_{a b}(\bq, \epsilon - i \eta, \epsilon + \omega + i \eta) \right. \nonumber \\
&+&\left. 2 i n_F(\epsilon) \textrm{Im} P_{a b} (\bq, \epsilon + i \eta, \epsilon + \omega + i \eta) \right]. 
\eeqa
Here $n_F(\epsilon)$ is the Fermi-Dirac distribution function, $\eta = 0+$, and $P = \gamma^{-2} \cI \cD$, where 
\beqa
\label{eq:5}
&&\cI_{a b}(\bq, \omega) = \frac{\gamma^2}{2} \tau^a_{\sigma_2 \sigma_1} \tau^b_{\sigma_3 \sigma_4} \int \frac{d^3 k}{(2 \pi)^3} \sum_{s s'} 
\nonumber \\ &\times&G^R_{\sigma_1 \sigma_3}(\bk + \bq, s,  \omega) G^A_{\sigma_4 \sigma_2}(\bk, s', 0), 
\eeqa 
and $\cD = (1 - \cI)^{-1}$ is the inverse diffusion propagator (diffuson). 
We will be interested in the long-wavelength low-frequency response, i.e. will evaluate $\chi(\bq,\omega)$ at small $q$ and $\omega$, 
such that $v_F q, \omega \ll 1/\tau, \epsilon_F, v _F Q$. 
The condition $q \ll Q$, in particular, means that the states, labelled by $\bk$ and $\bk + \bq$ in Eq.~\eqref{eq:5}, always belong to the same valley, if they are in the vicinity of the Fermi surface. 
This property is a consequence of the separation of the Weyl nodes in momentum space and would not exist in a Dirac semimetal, with bands 
weakly split by inversion symmetry breaking. As a result, the spin-charge coupling physics in the latter system would be completely different, 
similar to an ordinary metal or a doped semiconductor with weak spin-orbit coupling~\cite{Burkov04}. 
Conversely, the physics we will find in our system is specific to a chiral topological semimetal, with well-separated band-touching nodes. 

Evaluating the matrix $\cI$ at small $\omega$ and $q$, one obtains
\beqa
\label{eq:6}
&&\cD^{-1}(\bq, \omega) = \nonumber \\
&&\left(
\begin{array}{cc}
-i \omega \tau + D q^2 \tau & - i \Gamma q \tau \\
- i \Gamma q \tau &- \frac{i}{3} \omega \tau + \frac{2}{3} + \frac{3}{5} D q^2 \tau
\end{array}
\right), \nonumber \\
\eeqa
where $D = v_F^2 \tau/3$ is the diffusion constant and 
\beq
\label{eq:7} 
\Gamma = \frac{2 v_F \epsilon_F \Delta}{3 (\epsilon_F^2 + \Delta^2)}, 
\eeq
describes the coupling between the charge and the longitudinal component of the spin (along $\bq$), which correspond to the entries in the $2 \times 2$ matrix 
in Eq.~\eqref{eq:6}. The two transverse components of the spin, which do not couple to the charge density, have been neglected. 
The fact that the coefficient of $- i \omega \tau$ term in the second row of the inverse diffuson matrix is not $1$, but $1/3$, is a consequence 
of spin nonconservation due to the spin-momentum locking. Indeed, this coefficient is tied to the constant term $2/3$ (determining the spin relaxation rate) in the same matrix element, by an exact relation $2/3 = 1 - 1/3$. This, in turn, follows from the fact that any matrix element of the matrix $\cI(\bq, \omega)$, 
which does not vanish in the limit $q = 0$, has the Drude form $\cI(0, \omega) \sim 1/(1 - i \omega \tau)$, see the Appendix for details. 

We may view Eq.~\eqref{eq:6} as the inverse of the Fourier transformed Green's function of coupled spin-charge transport equations, which are given by
\beqa
\label{eq:8}
&&\frac{\partial n}{\partial t} = D \nabla^2 n + \Gamma \bnabla \cdot \bS, \nonumber \\
&&\frac{\partial S^a}{\partial t} = \frac{9}{5} D \nabla^2 S^a - \frac{2 S^a}{\tau} + 3 \Gamma \frac{\partial n}{\partial x_a}. 
\eeqa
Strictly speaking, such equations are valid on time and length scales, longer than $\tau$ and $\ell = v_F \tau$ (mean free path). 
The dynamics on significantly shorter scales is ballistic and 
is not described by simple diffusion equations, such as Eq.~\eqref{eq:8}. In what follows, we will assume that we may use these equations 
down to length scales of at least order $\ell$, expecting the diffusive-ballistic dynamics crossover to be smooth. 
 
Let us now solve these equations in the steady-state, assuming transport in the $x$-direction, in a sample that is uniform in the $y$ and $z$-directions. 
In this case, Eq.~\eqref{eq:8} becomes
\beqa
\label{eq:9}
&&\frac{d^2 n}{d x^2} + \frac{1}{L_{cs}} \frac{d S^x}{d x} = 0, \nonumber \\
&&\frac{d^2 S^x}{d x^2} - \frac{10 S^x}{3 \ell^2} + \frac{5}{3 L_{cs}} \frac{d n}{d x} = 0, 
\eeqa
where 
\beq
\label{eq:10}
L_{cs} = \frac{D}{\Gamma} = \ell \frac{\epsilon_F^2 + \Delta^2}{2 \epsilon_F \Delta}, 
\eeq
is a new length scale, that characterizes the strength of the spin-charge coupling. 
Note that, as defined, $L_{cs}$ may have either sign, determined by the sign of the product $\epsilon_F \Delta$. 
The coupling is strong when this length scale is short, i.e. of the order of the mean free path $\ell$. 
Since the charge is exactly conserved, we may identify
\beq
\label{eq:11}
j = e D \frac{d n}{d x} + e \Gamma S^x, 
\eeq
as the electric current density. 
Expressing $dn /dx$ in terms of $j$ and substituting in the second of Eqs.~\eqref{eq:9}, we obtain
\beq
\label{eq:12}
\frac{d^2 S^x}{d x^2} - \frac{S^x}{L_s^2} = - \frac{5 j}{e v_F \ell L_{cs}}, 
\eeq
where 
\beq
\label{eq:13}
L_s = \ell \sqrt{\frac{3/10}{1 + 2 \left(\frac{\epsilon_F \Delta}{\epsilon_F^2 + \Delta^2} \right)^2}}, 
\eeq
is the spin relaxation length.  
\begin{figure}
\centering
\includegraphics[width=\linewidth]{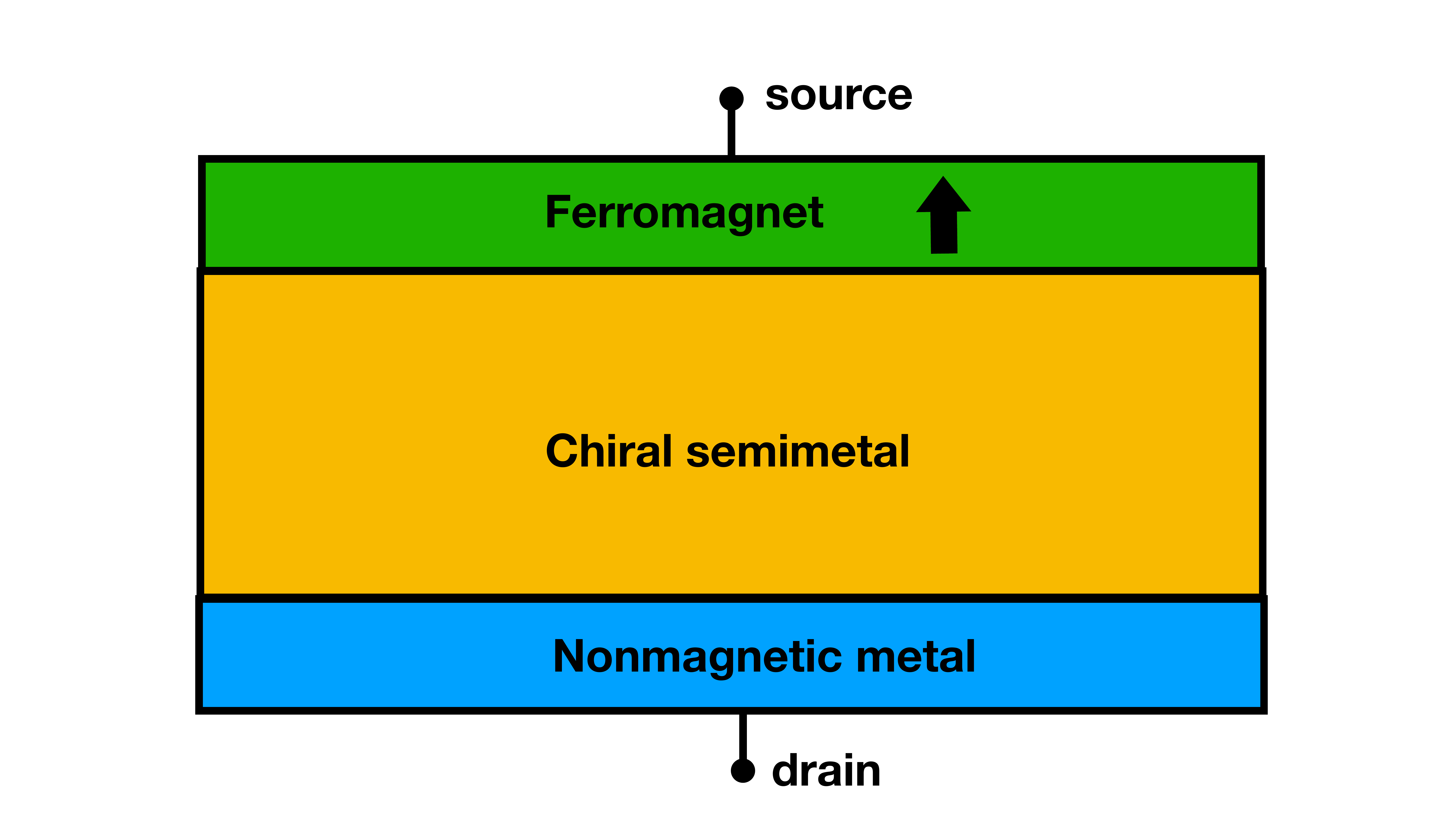}
\caption{(Color online) Cartoon of the magnetoresistance measurement setup. The current is injected from a polarized ferromagnetic electrode into the chiral semimetal sample and extracted by a normal metal electrode.}
\label{fig:3}
\end{figure}

Let us solve this equation in a semi-infinite sample, occupying the $x > 0$ half-space. 
We will assume that the electric current is injected into the sample at $x = 0$ from a ferromagnetic electrode, see Fig.~\ref{fig:3}. 
We will take the current to be spin-polarized in the $x$-direction, with the degree of polarization $\xi$. 
\begin{figure}[t]
\subfigure{
\includegraphics[width=\columnwidth]{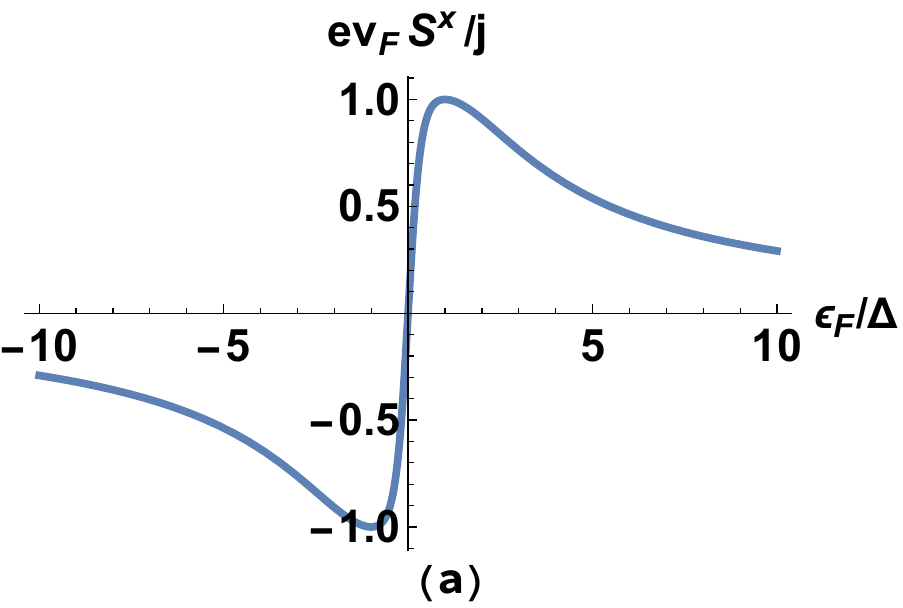}}
\subfigure{
\includegraphics[width=\columnwidth]{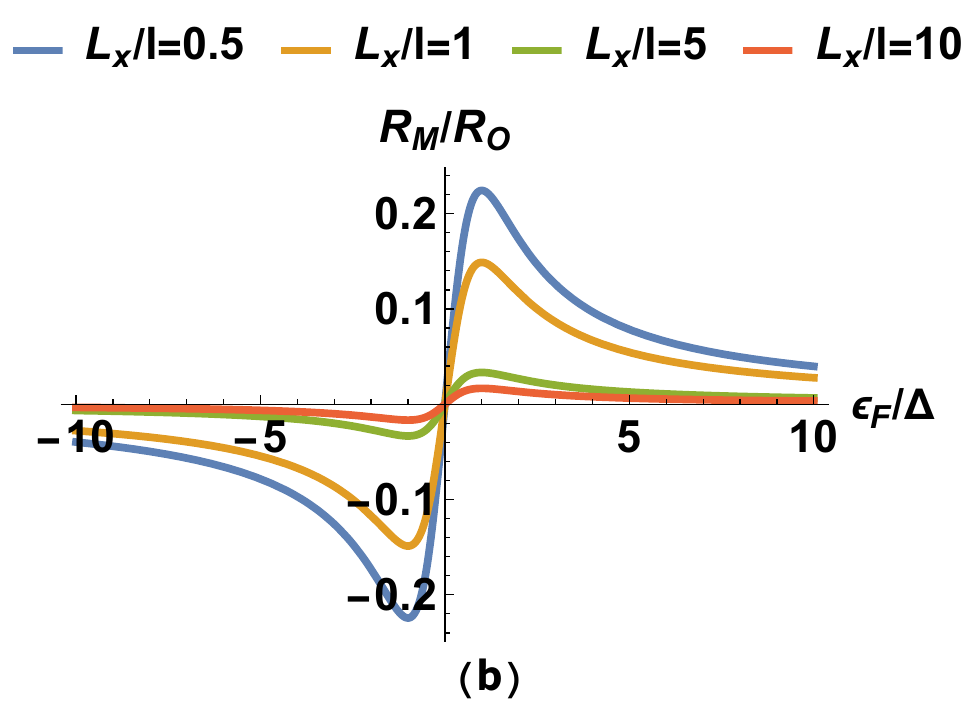}}
\caption{(Color online) Bulk current-induced spin accumulation [first term in Eq.~\eqref{eq:15}] (a) and magnetoresistance (b) as a function of the Fermi energy. Both signals vanish at charge neutrality and far away from the Weyl nodes and reach maximum magnitude when the Fermi energy coincides with the location of one of the nodes $\epsilon_F = \pm \Delta$. The maximum possible magnitude of the spin accumulation (i.e. $\hbar/2$ per electron) is normalized to unity.}
\label{fig:4}
\end{figure}
This corresponds to the boundary condition 
\beq
\label{eq:14}
\left. \frac{9 D}{5} \frac{d S^x}{d x} \right|_{x = 0} = \frac{j \xi}{e}. 
\eeq
Note that, naively, the last term in Eq.~\eqref{eq:8} also gives a contribution, $3 \Gamma n$, to the spin current. 
However, this contribution exists even in equilibrium and simply reflects the fact that states of opposite spin also have opposite 
momentum due to the spin-momentum locking near a Weyl node. Such equilibrium ``spin currents" do not correspond to actual transport 
of spin and thus should not be included into the boundary condition of Eq.~\eqref{eq:14}~\cite{Galitski06}. 

The solution of Eq.~\eqref{eq:12} with the boundary condition \eqref{eq:14} is 
\beq
\label{eq:15}
S^x(x) =  \frac{5 j L_s^2}{e v_F \ell L_{cs}} \left(1 - \frac{L_{cs} \xi}{3 L_s} e^{-x/ L_s} \right). 
\eeq
The first term in Eq.~\eqref{eq:15} is the intrinsic spin polarization of the charge current in the chiral semimetal, which arises due to the 
spin-momentum locking~\cite{Hueso22,Krieger22}. The second term is the extra spin density, injected from the ferromagnetic electrode. 
Substituting this back into Eq.~\eqref{eq:11}, we can find the voltage at a distance $L_x$ from the magnetic electrode in response to the injected current $I = j L_y L_z$
\beq
\label{eq:16}
V = \frac{1}{e g} \int_0^{L_x} \frac{d n}{d x} dx  = I (R_O + \xi R_M), 
\eeq
where 
\beq
\label{eq:17}
R_O = \frac{10 L_s^2 L_x}{e^2 g v_F \ell^3  L_y L_z}, 
\eeq
and 
\beq
\label{eq:18}
R_M = \frac{5 L_s^2}{3 e^2 g v_F \ell L_{cs} L_y L_z} \left(1 - e^{-L_x/L_s} \right). 
\eeq
Note that, we view $n/g$ in Eq.~\eqref{eq:16} as the full electrochemical potential, including the contribution of the external electrostatic potential, which has been
subsumed into $n/g$ for brevity. 

The first term in Eq.~\eqref{eq:16} corresponds to ohmic resistance, including the contribution of the nonequilibrium current-induced spin density, 
i.e. the first term in Eq.~\eqref{eq:15}. 
The second term is a nonohmic contribution with a nontrivial scale dependence, which arises from the extra spin polarization, injected 
from the ferromagnetic electrode. 
It is convenient to consider the ratio of the nonohmic and ohmic contributions, which may be used to characterize the strength of the magnetoresistance 
effect, arising from the spin-polarized current injection
\beq
\label{eq:19}
\frac{R_{M}}{R_O} = \frac{\ell \epsilon_F \Delta}{3 (\epsilon_F^2 + \Delta^2) L_x} \left(1 - e^{-L_x/L_s} \right). 
\eeq

Eq.~\eqref{eq:19}, along with Eq.~\eqref{eq:15} for the current-induced spin polarization, are the main results of this paper. 
One of the most important features of both is a nonmonotonic dependence on the Fermi energy. 
Both signals vanish at charge neutrality and when the Fermi energy is far away from the Weyl points, and reach maximum magnitude 
when the Fermi energy coincides with one of the Weyl points, i.e. when $\epsilon_F = \pm \Delta$, see Fig.~\ref{fig:4}.
This makes it clear that the effect originates from the Weyl nodes. 

Another characteristic feature of the magnetoresistance Eqs.~\eqref{eq:16},~\eqref{eq:19} is that it changes sign upon rotation of the magnetization of the 
ferromagnetic electrode by $\pi$. This is similar to a spin-valve effect, although in this case it does not require multiple magnetic layers. 
The overall magnitude of the magnetoresistance signal declines with the distance between the electrodes, although not very fast, as $1/L_x$ at long distances. 
To maximize the effect one would still want a sample in the form of a thin film, with thickness that is not significantly larger than the mean free path $\ell$, 
which should not pose a problem. 
\section{Conclusions}
\label{sec:3}
In conclusion, we have developed a theory of current-induced spin accumulation and magnetoresistance in chiral semimetals, in which Weyl nodes of opposite topological charge occur at different energies due to broken inversion and mirror symmetries. 
One of our main results is the demonstration that the current-induced spin accumulation is maximized when the Fermi energy coincides with the location
of one of the nodes, while vanishing both at charge neutrality and when the Fermi energy is far away from the nodes. The maximal spin accumulation, 
achieved in this case, is identical to that for an isolated Weyl node, which can only be realized as a surface state of a 4D quantum Hall insulator. 
 Our second result is the existence in this system of a unique magnetoresistance effect, closely related to the current-induced spin accumulation, wherein 
 nonequilibrium spin density, injected from a spin-polarized electrode, leads to an extra contribution to the electrical conductivity. 
 Its sign depends on the relative orientation of the spin polarization of the electrode and the current in a way that distinguishes it from most other 
 magnetoresistance mechanisms. In particular, it changes sign upon $\pi$-rotation of the magnetization. The magnitude of this effect is also a nonmonotonic 
 function of the Fermi energy, reaching maximum when the Fermi energy coincides with the location of one of the Weyl nodes. 
 This effect should be observable in thin films of currently known chiral semimetals, such as CoSi and RhSi, although many details will certainly be different
 due to a much more complex electronic structure of these materials, compared to our simple toy model. 
 These details will be investigated in future work. 
 
\begin{acknowledgments}
We acknowledge financial support from the Natural Sciences and Engineering Research Council (NSERC) of Canada. 
AAB, MS and IM were supported by Center for Advancement of Topological Semimetals, an Energy Frontier Research Center funded by the U.S. Department of Energy Office of Science, Office of Basic Energy Sciences, through the Ames Laboratory under
contract DE-AC02-07CH11358. 
Research at Perimeter Institute is supported in part by the Government of Canada through the Department of Innovation, Science and Economic Development and by the Province of Ontario through the Ministry of Economic Development, Job Creation and Trade.
\end{acknowledgments}

\begin{appendix}
\section{Calculation of the diffusion propagator}
Here we provide details of the calculations, omitted in the main text. 
We will start from the SCBA equation for the impurity self-energy in valley $s$ and band $r$, which is given by
\beq
\label{eq:A1}
\Sigma^R_{s r}(\bk, \omega) = \gamma^2 \int \frac{d^3 k'}{(2 \pi)^3} |\langle z^{s r}_{\bk} | z^{s' r'}_{\bk'} \rangle |^2 G^R_{s' r'}(\bk', \omega). 
\eeq
Here summation over repeated $s', r'$ indices is implied, 
\beq
\label{eq:A2}
G^R_{s r}(\bk, \omega) = \frac{1}{\omega - \xi_{s r}(\bk) - i \textrm{Im} \Sigma^R_{s r}(\bk, \omega)}, 
\eeq
is the retarded Green's function and the wavefunctions $|z^{s r}_\bk \rangle$ are given by
\beq
\label{eq:A3}
|z^{s r}_\bk \rangle = \left(\sqrt{\frac{1}{2} \left(1 + \frac{s r k_z}{k} \right)}, \frac{ s r k_+}{k} \sqrt{\frac{1}{2} \left(1 - \frac{s r k_z}{k} \right)}\right)^T. 
\eeq
Using 
\beq
\label{eq:A4}
\textrm{Im} G^R_{s r}(\bk, \omega) \approx - i \pi \delta[\xi_{s r}(\bk)], 
\eeq
we obtain
\beq
\label{eq:A5} 
\textrm{Im} \Sigma^R_{s r}(\bk, \omega) = - \frac{i \pi \gamma^2 g}{2} \equiv - \frac{i}{2 \tau}, 
\eeq
where 
\beq
\label{eq:A6}
g = \frac{1}{2 \pi^2 v_F^3}\left[(\epsilon_F - \Delta)^2 + (\epsilon_F + \Delta)^2 \right] = \frac{\epsilon_F^2 + \Delta^2}{\pi^2 v_F^3}, 
\eeq
is the density of states at Fermi energy. 

We now move on to the calculation of the diffusion propagator. 
We start from the general expression for the elements of the matrix $\cI$
\beqa
\label{eq:A7}
\cI_{a b}(\bq, \omega)&=&\frac{\gamma^2}{2} \tau^a_{\sigma_2 \sigma_1} \tau^b_{\sigma_3 \sigma_4} \int \frac{d^3 k}{(2 \pi)^3} \sum_{s s'} 
G^R_{\sigma_1 \sigma_3}(\bk + \bq, s,  \omega) \nonumber \\
&\times&G^A_{\sigma_4 \sigma_2}(\bk, s', 0).
\eeqa
When $q \ll Q$, which we assume, $\bk + \bq$ and $\bk$ in Eq.~\eqref{eq:7} always belong to the same valley, which gives
\beqa
\label{eq:A8}
&&\cI_{a b}(\bq, \omega) = \frac{\gamma^2}{2} \tau^a_{\sigma_2 \sigma_1} \tau^b_{\sigma_3 \sigma_4} \int \frac{d^3 k}{(2 \pi)^3} \sum_s
G^R_{\sigma_1 \sigma_3}(\bk + \bq, s,  \omega) \nonumber \\
&\times&G^A_{\sigma_4 \sigma_2}(\bk, s, 0) = \frac{\gamma^2}{2} \tau^a_{\sigma_2 \sigma_1} \tau^b_{\sigma_3 \sigma_4} \int \frac{d^3 k}{(2 \pi)^3} \sum_s \nonumber \\
&\times&\left[G^R_0(\bk, s, \omega) \tau^0_{\sigma_1 \sigma_3} + \bG_t^R(\bk + \bq, s,  \omega) \cdot \btau_{\sigma_1 \sigma_3}\right] \nonumber \\
&\times&\left[G^A_0(\bk, s, 0) \tau^0_{\sigma_4 \sigma_2} + \bG_t^A(\bk, s,  0) \cdot \btau_{\sigma_4 \sigma_2}\right]. 
\eeqa
This expression is a sum of pairwise products of retarded and advanced Green's functions, integrated over momentum. 
These are evaluated following the standard procedure. 
For example, consider
\beqa
\label{eq:A9}
&&\Gamma_{00}(\bq, \omega) \equiv \gamma^2 \int \frac{d^3 k}{(2 \pi)^3} \sum_s G^R_0(\bk + \bq, s, \omega) G^A_0(\bk, s, 0) \nonumber \\
&=&\frac{\gamma^2}{4} \int \frac{d^3 k}{(2 \pi)^3} \sum_{s r} \frac{1}{\omega - \xi_{s r}(\bk + \bq) + \frac{i}{2 \tau}} \frac{1}{0 - \xi_{s r}(\bk) - \frac{i}{2 \tau}}. \nonumber \\
\eeqa
Using the identity
\beq
\label{eq:A10}
A B = \frac{B - A}{A^{-1} - B^{-1}}, 
\eeq
we obtain
\beqa
\label{eq:A11}
&&\Gamma_{00}(\bq, \omega) \nonumber \\
&=&\frac{\gamma^2}{4} \int \frac{d^3 k}{(2 \pi)^3} \sum_{s r} \frac{1}{\omega - \xi_{s r}(\bq + \bq) + \xi_{s r}(\bk) + \frac{i}{\tau}} \nonumber \\
&\times&\left(\frac{1}{-\xi_{s r}(\bk) - \frac{i}{2 \tau}} - \frac{1}{\omega - \xi_{s r}(\bk + \bq) + \frac{i}{2 \tau}}\right). 
\eeqa
We now assume $\omega, v_F q, 1/\tau \ll |\epsilon_F \pm \Delta|$. This always breaks down when the Fermi energy approaches the location of a Weyl node, 
but, as mentioned in the main text, we will ignore this as all of the final results are smooth functions of $\epsilon_F$ and $\Delta$. 
Then we obtain
\beq
\label{eq:A12}
\Gamma_{00}(\bq, \omega) \approx \frac{\gamma^2}{4} \int \frac{d^3 k}{(2 \pi)^3} \sum_{s r} \frac{2 \pi i \delta[\xi_{s r}(\bk)]}{\omega - \bq \cdot \bnabla_\bk \xi_{s r}(\bk) 
+ \frac{i}{\tau}}. 
\eeq
What remains is thus a Fermi-surface integral, which is straightforward to evaluate. Then we finally obtain
\beq
\label{eq:A13}
\Gamma_{00}(\bq, \omega) = \frac{i}{4 q \ell} \ln\left(\frac{1 - i \omega \tau - i q \ell}{1 - i \omega \tau + i q \ell} \right), 
\eeq
where $\ell = v_F \tau$ is the mean free path. 
All other momentum integrals in Eq.~\eqref{eq:8} are evaluated in the same manner. 
Then we obtain the following results for the matrix elements of $\cI$
\beqa
\label{eq:A14}
&&\cI_{00}(\bq, \omega) = \frac{i}{2 q \ell} \ln\left(\frac{1 - i \omega \tau - i q \ell}{1 - i \omega \tau + i q \ell} \right), \nonumber \\
&&\cI_{0 \hat q}(\bq, \omega) = \cI_{\hat q 0}(\bq, \omega) =  \frac{2 i \epsilon_F \Delta}{q \ell (\epsilon_F^2 + \Delta^2)} \nonumber \\
&\times& \left[1 + \frac{1 - i \omega \tau}{2 i q \ell} 
\ln\left(\frac{1 - i \omega \tau - i q \ell}{1 - i \omega \tau + i q \ell} \right)\right], \nonumber \\
&&\cI_{\hat q \hat q}(\bq, \omega) = - \frac{1 - i \omega \tau}{(i q \ell)^2} \nonumber \\
&\times&\left[1 + \frac{1 - i \omega \tau}{2 i q \ell} \ln\left(\frac{1 - i \omega \tau - i q \ell}{1 - i \omega \tau + i q \ell} \right)\right].
\eeqa
Expanding these expressions to first order in $\omega \tau$ and up to second order in $q \ell$, we obtain Eq.~6 in the main text. 
These expressions may also be used in their original form to study diffusive-ballistic crossover phenomena when $\omega \tau$ and $q \ell$ may not be assumed to be small. 
One of course loses the simplicity and convenience of a partial differential equation (Eq.~8 in the main text) description in this regime. 
\end{appendix}

\bibliography{references}
\end{document}